\documentclass[aps,pra,twocolumn,superscriptaddress,nofootinbib,longbibliography]{revtex4-2}

\usepackage{amsmath,amssymb,graphicx,bm}
\usepackage[colorlinks=true,linkcolor=blue,citecolor=blue,urlcolor=blue]{hyperref}

\newcommand{\Fbar}{\bar{F}}
\newcommand{\Neg}{\mathcal{N}}

\begin{document}

\title{Teleportation through time-varying channels: threshold geometry and a complete-positivity bound on non-Markovian backflow}

\author{C. SEIDA}
\affiliation{Laboratoire InterDisciplinaire de Recherches Appliquées, LIDRA, Université Internationale d’Agadir – Universiapolis, Agadir, Morocco}

\date{\today}

\begin{abstract}
A Bell pair distributed through a link that combines amplitude damping with
dephasing at time-dependent rates has dynamics that separate into a fixed part
and a moving one. The negativity, fully entangled fraction, discord, and average
teleportation fidelity depend on time only through the accumulated damping
parameters $p(t)$ and $q(t)$. Every threshold is therefore a curve fixed in the
unit square, and the rates select nothing but a trajectory across it. The Horodecki fidelity formula $\bar{F} = \frac{1}{2} + \frac{1}{6}\mathrm{Tr}|T|$ covers
one- and two-sided exposure alike: the condition $\det T \leq 0$ under which it takes this
form holds throughout the unit square for both. Under symmetric
two-sided noise the entanglement vanishes when $p+q\geq1$, where the exact
relation $\bar{F}^{2s}=\frac{2}{3}+\frac{1}{3}\mathcal{N}_{2s}$ ties
disentanglement and the loss of quantum advantage to the same instant. One-sided
exposure admits no finite-time sudden death for any rate profile, and the discord
stays strictly positive throughout the open square, so the distributed state can
be separable, useless for teleportation, and still nonclassical. For harmonically
modulated rates, complete positivity caps the backflow at one modulation period of
static decay, $\Delta\Gamma_{k}\leq2\pi\gamma_{k,0}/\Omega$, equivalently at a
modulation depth $\xi_{k}\leq4.6033$ independent of $\Omega$. Inside that window
the trajectory reverses, producing finite intervals of restored quantum advantage
and entanglement sudden birth.
\end{abstract}

\maketitle

\section{Introduction}\label{sec:intro}

Quantum teleportation transfers an unknown qubit state using nothing more than a
shared entangled pair and two classical bits \cite{Bennett1993}, and that economy
is what makes it the elementary operation of a quantum network. Entanglement
swapping, remote gates and end-to-end state transfer all reduce to teleportation
acting on a distributed, and therefore imperfect, Bell pair
\cite{Kimble2008,Wehner2018,Azuma2023,Hu2023}. The primitive is now routine on
every major substrate: superconducting circuits joined by a cryogenic bus
\cite{Qiu2025,Chou2018}, photonically linked trapped-ion modules
\cite{Main2025}, solid-state spin registers
\cite{Hermans2022,Knaut2024,Wei2025}, and metropolitan and satellite optical
links \cite{Thomas2024,Liu2024,Rad2025,Ren2017}. These demonstrations share a
structure. A pair is distributed, held, and only then consumed. The resource
degrades in between, so the operational question is never whether entanglement
survives, but whether enough of it survives to beat the classical benchmark
\cite{Horodecki1999,RHorodecki1996}.

What degrades the resource is rarely a single mechanism. Energy relaxation at
rate $\gamma_{1}$ and pure dephasing at rate $\gamma_{\phi}$ act together, and at
static rates the entanglement of a distributed pair can vanish in finite time
while the local coherences decay only asymptotically
\cite{Yu2004,Yu2009,Almeida2007,Laurat2007}. The nonclassical correlations
measured by the quantum discord \cite{Ollivier2001,Henderson2001,Modi2012}
behave differently again, outliving the sudden death and fading only
asymptotically \cite{Werlang2009,Maziero2009,Mazzola2010,Ali2010,Chen2011}. No
single figure of merit therefore settles when a link has stopped being useful.

Real links are not stationary either. In superconducting circuits the relaxation
time $T_{1}$ and the dephasing time $T_{2}$ wander by factors of order unity over
minutes to hours as two-level-system defects drift through the qubit frequency
\cite{Klimov2018,Burnett2019,Schloer2019}, which is what motivated channel models
with explicitly time-dependent decoherence parameters
\cite{Martinez2021,Martinez2023}; atmospheric turbulence does the same to the
transmittance of a free-space link \cite{Vasylyev2016}. When an instantaneous
rate dips negative, the family of maps loses CP-divisibility and information
returns from the environment
\cite{,Breuer2009,Rivas2014,Breuer2016,Vega2017}, so that
entanglement which has already died can be reborn
\cite{Bellomo2007,Xu2010,LiuBH2011}.

Work on teleportation through such memory-bearing environments has concentrated
on producing fidelity revivals, through nonlocal memory effects
\cite{Laine2014}, weak measurement and its reversal \cite{Gaidi2026,Seida2021},
non-Markovian channel design \cite{Yeo2010,Wang2023,Zhang2024}, and directly in
experiment \cite{LiuZD2020}. What stays implicit in those treatments is a
separation that the composite channel makes exact. Negativity, concurrence,
fully entangled fraction, discord and average fidelity all depend on time only
through the two accumulated damping parameters $p(t)$ and $q(t)$. Every threshold
is therefore a curve fixed in the unit square, and a time-varying environment
moves none of them. It selects a path through a static landscape, nothing more.

This work makes that split explicit and addresses the two quantitative questions
it raises: how far back along a trajectory an environment is able to push the
state, and whether the resource is lost first by disentangling or by falling
below the classical benchmark. The Horodecki fidelity formula $\bar{F} = \frac{1}{2} + \frac{1}{6}\mathrm{Tr}|T|$
\cite{RHorodecki1996} applies to both exposure geometries without modification.
Within that landscape we obtain the sudden-death criterion $p(t) + q(t) \geq 1$
with the exact relation between fidelity and negativity that holds up to it, the absence of
any finite-time sudden death under one-sided exposure, a characterisation of the
useless-entanglement region \cite{Badziag2000,Verstraete2003} as an artifact of
one-sided decoherence, and a complete-positivity bound
$\Delta\Gamma_k \leq \gamma_{k,0} T_{\mathrm{mod}}$ on the attainable backflow.

The paper is organised as follows. Section~\ref{sec:model} sets up the model, the
time-varying amplitude- and phase-damping channels, and the complete-positivity
bound constraining non-Markovian backflow. Section \ref{sec:geometry} recasts the resource thresholds as a static geometry,
applying the fidelity formula to both exposure geometries and assembling a unified
threshold diagram whose boundaries separate entanglement, discord and the classical
benchmark. Section~\ref{sec:dynamics} traces static, fast-modulated and
modulated channels through that geometry, where entanglement sudden birth and the
ordering of the two lifetimes appear as trajectory crossings. 
Section~\ref{sec:conclusion} concludes.
\section{Model}
\label{sec:model}

\subsection{Figures of merit}

We use the standard protocol of Ref.~\cite{Bennett1993}, with a shared two-qubit
resource $\rho_{23}$ in place of the ideal pair
$|\Phi^{+}\rangle=(|00\rangle+|11\rangle)/\sqrt{2}$, and no preprocessing of the
resource. The state fidelity $F(\psi)=\langle\psi|\rho_{\rm out}|\psi\rangle$,
averaged uniformly over the Bloch sphere, obeys the Horodecki relation
\cite{Horodecki1999, RHorodecki1996}
\begin{equation}
\bar{F}=\frac{2f+1}{3},
\qquad
f=\max_{|e\rangle}\langle e|\rho_{23}|e\rangle ,
\label{eq:horodecki}
\end{equation}
where $f$ is the fully entangled fraction and $|e\rangle$ runs over all maximally
entangled two-qubit states. The classical benchmark $\bar{F}_{\rm cl}=2/3$
corresponds to $f=1/2$. Entanglement is measured by the negativity
$\mathcal{N}(\rho)=2\sum_{\lambda_{i}<0}|\lambda_{i}|$, where the $\lambda_{i}$
are the eigenvalues of the partial transpose $\rho^{T_{3}}_{23}$
\cite{Peres1996,Vidal2002}, normalised so that $\mathcal{N}=1$ for a maximally
entangled Bell state. Correlations that survive the loss of entanglement are
measured by the quantum discord \cite{Ollivier2001}, defined in
Sec.~\ref{sec:discord}.

\subsection{Time-varying amplitude damping and dephasing}

The link is the composition of an amplitude-damping (AD) and a phase-damping (PD)
channel with time-dependent rates, generated by the time-local master equation
\cite{Hall2014}
\begin{equation}
\frac{d\rho}{dt}
=\gamma_{1}(t)\Big[\sigma_{-}\rho\sigma_{+}
-\tfrac{1}{2}\{\sigma_{+}\sigma_{-},\rho\}\Big]
+\frac{\gamma_{\phi}(t)}{2}\Big[\sigma_{z}\rho\sigma_{z}-\rho\Big].
\label{eq:master}
\end{equation}
Integrating Eq.~(\ref{eq:master}) gives the two accumulated damping parameters
\begin{equation}
p(t)=1-e^{-\Gamma_{1}(t)},
\qquad
q(t)=1-e^{-2\Gamma_{\phi}(t)},
\label{eq:pq}
\end{equation}
with $\Gamma_{1}(t)=\int_{0}^{t}\gamma_{1}(s)\,ds$ and
$\Gamma_{\phi}(t)=\int_{0}^{t}\gamma_{\phi}(s)\,ds$. These two numbers, and not
the rates themselves, are the only time-dependent inputs to everything that
follows.

The AD and PD parts of Eq.~(\ref{eq:master}) commute, so the resulting map is the
composition of the two elementary channels and acts on an arbitrary qubit state
as
\begin{equation}
\Lambda_{t}[\rho]=\sum_{\mu=1}^{3}K_{\mu}(t)\,\rho\,K_{\mu}^{\dagger}(t),
\qquad
\sum_{\mu}K_{\mu}^{\dagger}K_{\mu}=\mathbb{I},
\label{eq:channel}
\end{equation}
with the three Kraus operators
\begin{eqnarray}
K_{1}&=&\begin{pmatrix}1&0\\[2pt]0&\sqrt{(1-p)(1-q)}\end{pmatrix},
\quad
K_{2}=\begin{pmatrix}0&\sqrt{p}\\[2pt]0&0\end{pmatrix},
\quad \\ \nonumber
K_{3}&=&\begin{pmatrix}0&0\\[2pt]0&\sqrt{q(1-p)}\end{pmatrix}.
\label{eq:kraus}
\end{eqnarray}
Equation~(\ref{eq:kraus}) is the set of products $B_{\mu}A_{\nu}$ of the standard
AD and PD Kraus operators
\begin{eqnarray}
A_{0}=\begin{pmatrix}1&0\\[2pt]0&\sqrt{1-p}\end{pmatrix},
\quad
A_{1}=\begin{pmatrix}0&\sqrt{p}\\[2pt]0&0\end{pmatrix},
\quad \\ \nonumber
B_{0}=\begin{pmatrix}1&0\\[2pt]0&\sqrt{1-q}\end{pmatrix},
\quad
B_{1}=\begin{pmatrix}0&0\\[2pt]0&\sqrt{q}\end{pmatrix},
\label{eq:elementary}
\end{eqnarray}
of Refs.~\cite{Nielsen2010,Breuer2002}, of which $B_{1}A_{1}=0$ identically,
leaving three. The time-dependent parametrisation~Eq. (\ref{eq:pq}) of these
operators, and the construction of a Kraus decomposition directly from a
time-local generator, follow Refs.~\cite{Martinez2021,Andersson2007}.

Because the completeness relation in Eq.~(\ref{eq:channel}) holds at every
instant, $\{\Lambda_{t}\}_{t\geq0}$ is CPTP whenever $p,q\in[0,1]$, that is
whenever $\Gamma_{1}(t),\Gamma_{\phi}(t)\geq0$.
Evaluating Eq.~(\ref{eq:channel}) in the
computational basis,
\begin{equation}
\Lambda_{t}
\begin{pmatrix}\rho_{00}&\rho_{01}\\[2pt]\rho_{10}&\rho_{11}\end{pmatrix}
=\begin{pmatrix}
\rho_{00}+p\,\rho_{11} & \eta\,\rho_{01}\\[2pt]
\eta\,\rho_{10} & (1-p)\rho_{11}
\end{pmatrix},
\label{eq:action}
\end{equation}
which exhibits a single coherence factor,
\begin{equation}
\eta(t)=\sqrt{\bigl(1-p(t)\bigr)\bigl(1-q(t)\bigr)}
=\exp\!\Big[-\tfrac{1}{2}\Gamma_{1}(t)-\Gamma_{\phi}(t)\Big].
\label{eq:eta}
\end{equation}
The map covers pure amplitude damping ($q\to0$), pure dephasing ($p\to0$),
complete relaxation ($p\to1$) and the identity ($p=q=0$) as limits.

Two exposure geometries occur in practice. In the first, one arm of the pair is
held in a memory good enough to be treated as noiseless while the other is
transmitted, so that only the second qubit is damped during distribution. The
resource is $\rho_{23}=(\mathbb{I}\otimes\Lambda_{t})[|\Phi^{+}\rangle\langle
\Phi^{+}|]$,
\begin{equation}
\rho_{23}=\tfrac{1}{2}\Big[|00\rangle\langle00|
+\eta\big(|00\rangle\langle11|+{\rm h.c.}\big)
+p\,|10\rangle\langle10|+(1-p)|11\rangle\langle11|\Big].
\label{eq:rho1s}
\end{equation}
In the second, both qubits are stored or sent through nominally identical links,
the map is $\Lambda_{t}\otimes\Lambda_{t}$, and
\begin{equation}
\rho^{2s}_{23}=\frac{1}{2}
\begin{pmatrix}
1+p^{2} & 0 & 0 & \eta^{2}\\
0 & p(1-p) & 0 & 0\\
0 & 0 & p(1-p) & 0\\
\eta^{2} & 0 & 0 & (1-p)^{2}
\end{pmatrix}.
\label{eq:rho2s}
\end{equation}
Both are $X$ states of unit trace for all $p,q\in[0,1]$. 

\subsection{Modulated rates and the complete-positivity bound on backflow}
\label{sec:backflow}
\begin{figure}[t]
\includegraphics[width=\columnwidth]{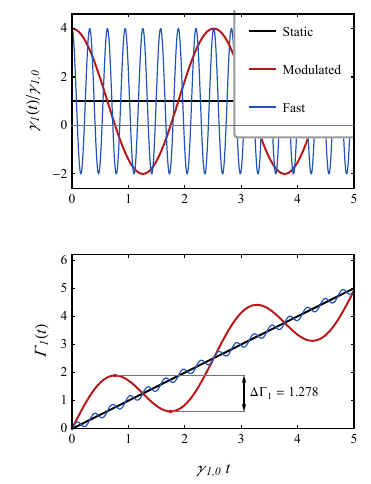}
\caption{Instantaneous rate $\gamma_{1}(t)$ (upper panel) and accumulated rate
$\Gamma_{1}(t)$ (lower panel) for the three regimes of
Sec.~\ref{sec:backflow}: static ($\xi=0$), modulated ($\xi=3$,
$\Omega=2.5\gamma_{1,0}$) and fast ($\xi=3$, $\Omega=20\gamma_{1,0}$). In both
modulated cases the instantaneous rate turns negative over part of each period,
but only in the modulated one does this leave a visible mark on
$\Gamma_{1}$: the accumulated rate peaks at $\gamma_{1,0}t\simeq0.76$ and
bottoms out at $\gamma_{1,0}t\simeq1.75$, separated by the peak-to-trough
excursion $\Delta\Gamma_{1}=1.278$ of Eq.~(\ref{eq:backflow}). Under fast
modulation the same depth produces an excursion of order
$\xi\gamma_{1,0}/\Omega$ and the curve is indistinguishable from the static one.
The mean rate is $\gamma_{1,0}$ in all three cases.}
\label{fig:rates}
\end{figure}
We now specialise to a concrete rate profile in order to quantify how much
decoherence an environment of this class is able to hand back. Any periodic
modulation admits a Fourier decomposition, and we retain the fundamental
harmonic,
\begin{equation}
\gamma_{k}(t)=\gamma_{k,0}\big[1+\xi_{k}\cos(\Omega t)\big],
\qquad k\in\{1,\phi\},
\label{eq:rate}
\end{equation}
with mean rate $\gamma_{k,0}$, modulation frequency $\Omega$ and dimensionless
depth $\xi_{k}$. The mean is preserved for every $\xi_{k}$, so the three regimes
below differ only in how the same total decoherence is delivered in time. The
accumulated rate stays in closed form,
\begin{equation}
\Gamma_{k}(t)=\gamma_{k,0}\Big[t+\frac{\xi_{k}}{\Omega}\sin(\Omega t)\Big].
\label{eq:Gamma}
\end{equation}
Periodic modulation of the coupling between a qubit and its environment is the
standard framework of dynamical decoherence control \cite{Kofman2001}, and
periodically driven open qubits are known to exhibit negative canonical rates and
controllable memory effects \cite{Amati2024,Follia2026}.

Figure~\ref{fig:rates} shows the instantaneous rate $\gamma_{1}(t)$ and the
accumulated rate $\Gamma_{1}(t)$ for the three regimes used throughout the paper.
For static rates ($\xi_{k}=0$) every resource measure decays monotonically. Under
fast modulation ($\Omega\gg\gamma_{k,0}$) the oscillatory term in
Eq.~(\ref{eq:Gamma}) is $\mathcal{O}(\xi_{k}\gamma_{k,0}/\Omega)$, and the lower
panel of Fig.~\ref{fig:rates} shows the accumulated rate tracking the static one
so closely that the two are indistinguishable; the channel collapses back onto
the static case. In the modulated regime, with $\Omega$ comparable to
$\gamma_{k,0}$ and $\xi_{k}>1$, the rate turns transiently negative, $\Gamma_{k}$
becomes non-monotonic, and the trajectory in the $(p,q)$ plane reverses. The
excursion visible between the peak at $\gamma_{1,0}t\simeq0.76$ and the trough at
$\gamma_{1,0}t\simeq1.75$ in Fig.~\ref{fig:rates} is the quantity we now bound.

\paragraph*{Backflow amplitude.}
The stationary points of $\Gamma_{k}$ satisfy $\cos\Omega t=-1/\xi_{k}$ and exist
only for $\xi_{k}>1$. Writing $a\equiv\arccos(-1/\xi_{k})$, the first peak lies at
$\Omega t=a$ and the first trough at $\Omega t=2\pi-a$, where
\begin{eqnarray}
\Gamma_{k}(t_{\rm peak})=&\frac{\gamma_{k,0}}{\Omega}
\Big[a+\sqrt{\xi_{k}^{2}-1}\Big],
\\ \nonumber
\Gamma_{k}(t_{\rm trough})=&\gamma_{k,0}T_{\rm mod}-\Gamma_{k}(t_{\rm peak}),
\label{eq:peaktrough}
\end{eqnarray}
with $T_{\rm mod}=2\pi/\Omega$. The peak-to-trough excursion, the decoherence
handed back to the system during one memory episode, is then
\begin{equation}
\Delta\Gamma_{k}=\frac{2\gamma_{k,0}}{\Omega}
\Big[\sqrt{\xi_{k}^{2}-1}+\arccos\big(-1/\xi_{k}\big)-\pi\Big],
\label{eq:backflow}
\end{equation}
which vanishes for $\xi_{k}\leq1$, grows with $\xi_{k}$ and decays as $1/\Omega$,
so that all three regimes follow from a single expression. For the modulated case
of Fig.~\ref{fig:rates}, $\xi=3$ and $\Omega=2.5\gamma_{1,0}$, it gives
$\Delta\Gamma_{1}=1.278$.

\paragraph*{The bound.}
Complete positivity requires $p,q\in[0,1]$, that is $\Gamma_{k}(t)\geq0$
throughout. The minima of Eq.~(\ref{eq:Gamma}) lie at $\Omega t=2\pi-a+2\pi n$ and
are ordered in $n$, so it suffices to impose positivity at the first one. By
Eq.~(\ref{eq:peaktrough}) that condition reads
\begin{equation}
\sqrt{\xi_{k}^{2}-1}+\arccos(-1/\xi_{k})\;\leq\;2\pi ,
\label{eq:cp}
\end{equation}
or equivalently $\Gamma_{k}(t_{\rm peak})\leq\gamma_{k,0}T_{\rm mod}$. Combining
the two expressions in Eq.~(\ref{eq:peaktrough}) gives the exact relation
\begin{equation}
\Delta\Gamma_{k}=2\,\Gamma_{k}(t_{\rm peak})-\gamma_{k,0}T_{\rm mod},
\label{eq:identity}
\end{equation}
so that the trough condition, the peak condition and the depth
condition~(\ref{eq:cp}) are one and the same, and
\begin{equation}
\Delta\Gamma_{k}\;\leq\;\Gamma_{k}(t_{\rm peak})\;\leq\;\gamma_{k,0}T_{\rm mod}.
\label{eq:bound}
\end{equation}
The environment can return no more decoherence than it has already deposited,
which holds for any rate profile, and for harmonic modulation the deposit itself
cannot exceed one modulation period of static decay. The second inequality is the
one that uses Eq.~(\ref{eq:rate}): a sharper waveform, depositing in a small
fraction of the period, evades it while still respecting the first. Since
$dS/d\xi_{k}=\sqrt{\xi_{k}^{2}-1}/\xi_{k}>0$ with $S(1)=\pi$, where $S$ denotes
the left-hand side of Eq.~(\ref{eq:cp}), that inequality has the unique root
$\xi_{\max}=4.6033$, and the admissible non-Markovian window
$1<\xi_{k}\leq4.6033$ is independent of $\Omega$.

One consequence of Eq.~(\ref{eq:bound}) is worth recording here, since it governs
everything in Sec.~\ref{sec:dynamics}. The drift of $\Gamma_{k}$ over one
modulation period is $\gamma_{k,0}T_{\rm mod}$, exactly the quantity that bounds
the fold. A harmonically modulated link therefore cannot push the trajectory back
past where it stood one period earlier, and each fixed threshold in the $(p,q)$
plane is re-crossed at most once, at any $\Omega$ and any admissible depth. At
$\xi_{k}=\xi_{\max}$ the fold equals the drift and the bound is saturated.
\section{The static threshold landscape}
\label{sec:geometry}

\subsection{Fidelity formula}

Both degraded states, Eqs.~(\ref{eq:rho1s}) and (\ref{eq:rho2s}), have a diagonal
correlation matrix $T_{ij}=\mathrm{Tr}[\rho\,\sigma_{i}\otimes\sigma_{j}]$,
\begin{equation}
T^{1s}=\begin{pmatrix}\eta&0&0\\0&-\eta&0\\0&0&1-p\end{pmatrix},
\quad
T^{2s}=\begin{pmatrix}\eta^{2}&0&0\\0&-\eta^{2}&0\\0&0&1-2p+2p^{2}\end{pmatrix}.
\label{eq:Tmatrices}
\end{equation}
Any maximally entangled two-qubit state has vanishing local Bloch vectors and an
orthogonal correlation matrix of determinant $-1$, so its overlap with $\rho$ is
$\frac{1}{4}[1+\mathrm{Tr}(TW^{\top})]$ with $W\in O(3)$, $\det W=-1$.
Maximising over $W$ gives $f=\frac{1}{4}[1+\sum_{i}s_{i}]$ in terms of the
singular values $s_{i}$ of $T$, valid whenever $\det T\leq0$, which holds for both
states here. With Eq.~(\ref{eq:horodecki}) this yields one expression
covering both geometries \cite{RHorodecki1996}
\begin{equation}
f=\frac{1+\mathrm{Tr}|T|}{4},
\qquad
\bar{F}=\frac{1}{2}+\frac{\mathrm{Tr}|T|}{6},
\label{eq:fidelity}
\end{equation}
so that the quantum-advantage condition $\bar{F}>2/3$ reads simply
$\mathrm{Tr}|T|>1$. Substituting Eq.~(\ref{eq:Tmatrices}),
\begin{align}
\bar{F}^{1s}&=\frac{4-p+2\eta}{6}, & &2\eta>p,
\label{eq:F1s}\\[2pt]
\bar{F}^{2s}&=\frac{2-p+p^{2}+\eta^{2}}{3}, & &\eta^{2}>p(1-p),
\label{eq:F2s}
\end{align}
where the right-hand inequalities are the corresponding advantage conditions. In
both cases the maximum in Eq.~(\ref{eq:horodecki}) is attained at
$|\Phi^{+}\rangle$. Written in the damping parameters, the two advantage
boundaries are the fixed curves
\begin{equation}
\underbrace{4(1-p)(1-q)=p^{2}}_{\text{one-sided}},
\qquad
\underbrace{p+q=1}_{\text{two-sided}},
\label{eq:boundaries}
\end{equation}
the second following from $\eta^{2}=(1-p)(1-q)$ after cancelling $1-p$, which
requires $p<1$.

\subsection{Entanglement}
\begin{itemize}
\item{One-sided exposure}
\end{itemize}

The partial transpose of Eq.~(\ref{eq:rho1s}) with respect to qubit 3 is block
diagonal.  So
\begin{equation}
\mathcal{N}(t)=2|\lambda_{-}|
=\frac{\sqrt{p^{2}(t)+4\eta^{2}(t)}-p(t)}{2}.
\label{eq:neg1s}
\end{equation}
One-sided exposure never produces
finite-time entanglement sudden death, whatever the time dependence of the rates.
The advantage condition $2\eta>p$ does fail at finite time, and the gap between
the two is the useless-entanglement region \cite{Badziag2000,Verstraete2003}
analysed in Sec.~\ref{sec:diagram}. 
\begin{figure}[t]
\includegraphics[width=\columnwidth]{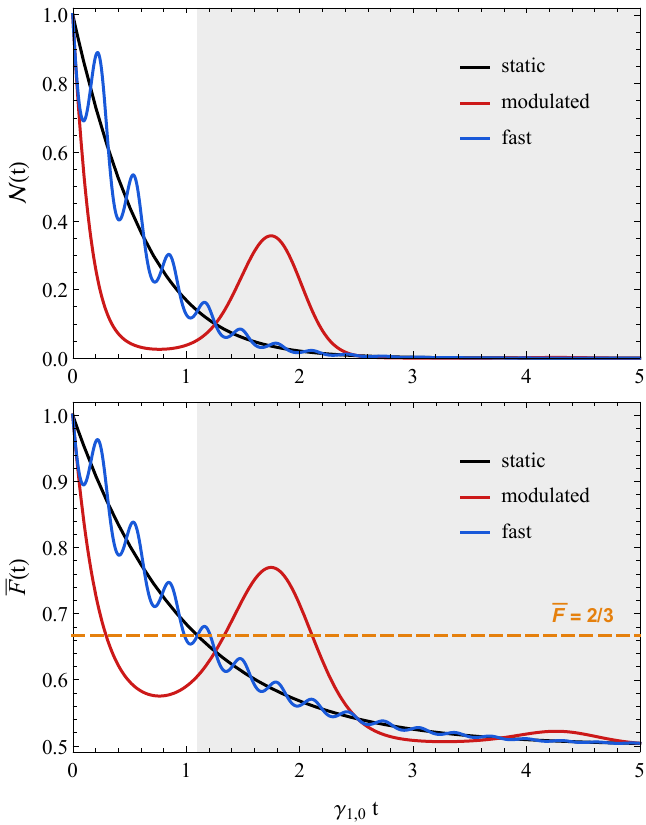}
\caption{One-sided exposure: negativity $\Neg(t)$ and average teleportation fidelity
$\Fbar(t)$ for the three regimes. The dashed line is the classical benchmark $\Fbar=2/3$ and
the shaded band is the useless-entanglement regime. Modulation drives excursions back above
the benchmark, fragmenting that regime into finite windows, while $\Neg$ never reaches zero.}
\label{fig:onesided}
\end{figure}

Figure~\ref{fig:onesided} shows the two
quantities together for the rate profiles of Sec.~\ref{sec:backflow}: the
negativity in the upper panel stays positive throughout, while the fidelity in
the lower panel crosses the benchmark and the shaded band opens.

\paragraph*{Two-sided exposure.}
The partial transpose of Eq.~(\ref{eq:rho2s}) has eigenvalues
$\frac{1}{2}[p(1-p)\pm\eta^{2}]$ in the $\{|01\rangle,|10\rangle\}$ block, whence
\begin{equation}
\mathcal{N}_{2s}(t)=\max\Big\{0,\;\eta^{2}(t)-p(t)\big[1-p(t)\big]\Big\},
\label{eq:neg2s}
\end{equation}
again unity at $p=q=0$. Using $\eta^{2}=(1-p)(1-q)$ and cancelling $1-p$, the
entanglement vanishes precisely when
\begin{equation}
p(t)+q(t)\;\geq\;1,
\label{eq:esd}
\end{equation}
which is Eq.~(\ref{eq:boundaries}) again. Two limits check the criterion. For
pure amplitude damping ($q=0$) it demands $p=1$, so a maximally entangled pair
under two-sided relaxation decays asymptotically without sudden death, in
agreement with Yu and Eberly \cite{Yu2004}. For complete dephasing ($q\to1$) it
is met immediately.

Comparing Eq.~(\ref{eq:neg2s}) with Eq.~(\ref{eq:F2s}) gives an exact linear
relation,
\begin{equation}
\bar{F}^{2s}(t)=\frac{2}{3}+\frac{1}{3}\,\mathcal{N}_{2s}(t),
\qquad \mathcal{N}_{2s}>0,
\label{eq:FN}
\end{equation}
\begin{figure}[t]
\includegraphics[width=\columnwidth]{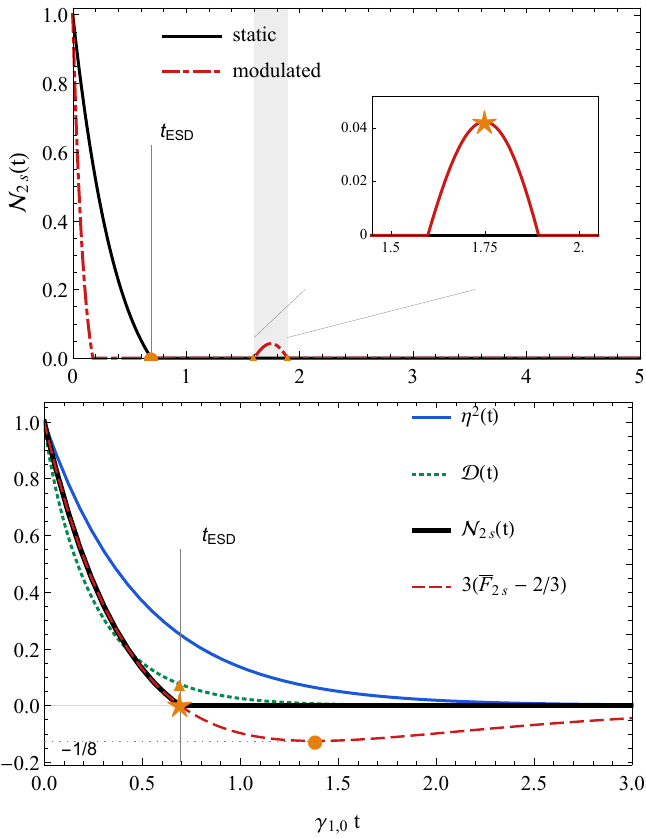}
\caption{Two-sided exposure. Upper panel: negativity $\Neg_{2s}(t)$ for the static and
modulated channels; the static channel dies once at $t_{\rm ESD}=\ln2$, while modulation
produces a single death--birth--death cycle (shaded) at $\gamma_{1,0}t\simeq0.178$, $1.599$,
$1.893$. Lower panel: ordering of the static lifetimes, normalised to unity at $t=0$:
$\eta^2(t)$, $\mathcal{D}(t)$, $2\Neg_{2s}(t)$ and the fidelity margin
$3(\Fbar^{2s}-2/3)$. The last two coincide for $t<t_{\rm ESD}$ by Eq.~\eqref{eq:FN}; past
$t_{\rm ESD}$ the negativity is clamped at zero while the margin dips to $-1/8$ at
$\gamma_{1,0}t=\ln4$.}
\label{fig:twosided}
\end{figure}
so under symmetric two-sided noise the negativity and the teleportation fidelity
are in one-to-one correspondence, and sudden death coincides exactly with the
loss of quantum advantage. The upper panel of Fig.~\ref{fig:twosided} shows the
negativity vanishing at a finite time, in contrast with
Fig.~\ref{fig:onesided}.
Nothing similar holds one-sided. There Eqs.~(\ref{eq:neg1s}) and (\ref{eq:F1s})
have different zero sets, and the gap between them is exactly the
useless-entanglement region. Equation~(\ref{eq:FN}) relies on the two arms being
statistically identical. If they are not, as when a memory qubit and a
transmitted qubit see different $(p,q)$, the correlation matrix retains the
symmetry $c_{2}=-c_{1}$, with $c_{1}$ now the product of the two coherence
factors, so Eq.~(\ref{eq:fidelity}) continues to apply. Whether the
fidelity-negativity relation itself survives is a quantitative question we leave
open.

\subsection{Quantum discord}
\label{sec:discord}

Correlations that outlive the entanglement are measured by the quantum discord
\cite{Ollivier2001,Henderson2001}. Taking the measurement on qubit 2, the arm
Alice retains,
\begin{equation}
\mathcal{D}=S(\rho_{2})-S(\rho_{23})+\min_{\hat{n}}\tilde{S}(\hat{n}),
\qquad
\tilde{S}(\hat{n})=\sum_{\pm}p_{\pm}\,S\big(\rho_{3|\pm}\big),
\label{eq:discord}
\end{equation}
where $\rho_{3|\pm}$ are the conditional states of qubit 3 produced by the
projectors $\Pi_{\pm}=\frac{1}{2}(\mathbb{I}\pm\hat{n}\cdot\vec{\sigma})$. Both
degraded states take the Bloch form
\begin{equation}
\rho=\frac{1}{4}\Big[\mathbb{I}\otimes\mathbb{I}+a\,\sigma_{z}\otimes\mathbb{I}
+b\,\mathbb{I}\otimes\sigma_{z}+\sum_{i}c_{i}\,\sigma_{i}\otimes\sigma_{i}\Big],
\label{eq:bloch}
\end{equation}
with $c_{2}=-c_{1}$ and
\begin{equation}
(a,b,c_{1},c_{3})=
\begin{cases}
\big(0,\,p,\,\eta,\,1-p\big), & \text{one-sided},\\[4pt]
\big(p,\,p,\,\eta^{2},\,1-2p+2p^{2}\big), & \text{two-sided},
\end{cases}
\label{eq:blochparams}
\end{equation}
so that $S(\rho_{2})=1$ and $h[(1+p)/2]$ respectively, with
$h(x)=-x\log_{2}x-(1-x)\log_{2}(1-x)$. Like every other figure of merit here,
$\mathcal{D}$ depends on time only through $p$ and $q$.

A measurement along $\hat{n}(\theta,\varphi)$ occurs with probability
$p_{\pm}=(1\pm a\cos\theta)/2$ and leaves qubit 3 with Bloch length
\begin{equation}
\kappa_{\pm}(\theta)=
\frac{\sqrt{c_{1}^{2}\sin^{2}\theta+(b\pm c_{3}\cos\theta)^{2}}}
{1\pm a\cos\theta},
\label{eq:kappa}
\end{equation}
independent of $\varphi$ because $|c_{1}|=|c_{2}|$, so that
$\tilde{S}(\theta)=\sum_{\pm}p_{\pm}h[(1+\kappa_{\pm})/2]$. The replacement
$\theta\to\pi-\theta$ exchanges $p_{\pm}$ with $p_{\mp}$ and $\kappa_{\pm}$ with
$\kappa_{\mp}$, leaving $\tilde{S}$ unchanged, so the minimisation runs over
$\theta\in[0,\pi/2]$ alone. Throughout the unit square the minimum sits at an
endpoint,\footnote{Restricting the optimisation to $\theta=0$ and $\theta=\pi/2$
is the usual $X$-state prescription, known to fail on a measure-zero set
\cite{Chen2011}. A scan of $10^{3}$ values of $\theta$ on a grid covering the
square returned an endpoint minimum at every point tested.} and both endpoints
are realised in different regions:
\begin{eqnarray}
\tilde{S}^{1s}_{\parallel}&=&\tfrac{1}{2}h(p),
\quad
\tilde{S}^{1s}_{\perp}=h\Big[\tfrac{1}{2}\big(1+\sqrt{p^{2}+\eta^{2}}\big)\Big],
\label{eq:S1s}\\ \nonumber
\tilde{S}^{2s}_{\parallel}&=&\frac{1+p}{2}\,h\!\left(\frac{1+p^{2}}{1+p}\right)
+\frac{1-p}{2}\,h(p), \\ \nonumber
\tilde{S}^{2s}_{\perp}&=&h\Big[\tfrac{1}{2}\big(1+\sqrt{p^{2}+\eta^{4}}\big)\Big].
\label{eq:S2s}
\end{eqnarray}
The longitudinal branch depends on $p$ alone in both geometries. The joint
entropies follow from the $X$ structure,
\begin{eqnarray}
S(\rho_{23})&=&-\!\!\sum_{\lambda\in\{\lambda_{+},\lambda_{-},p/2\}}\!\!
\lambda\log_{2}\lambda,
\\ \nonumber
\lambda_{\pm}&=&\tfrac{1}{4}\big[(2-p)\pm R\big],
\label{eq:S1sjoint}
\end{eqnarray}
\begin{eqnarray}
S(\rho^{2s}_{23})&=-\!\!\sum_{\mu\in\{\mu_{+},\mu_{-}\}}\!\!\mu\log_{2}\mu
-p(1-p)\log_{2}\frac{p(1-p)}{2},
\\ \nonumber
\mu_{\pm}=&\tfrac{1}{2}\Big[1-p+p^{2}\pm\sqrt{p^{2}+\eta^{4}}\Big],
\label{eq:S2sjoint}
\end{eqnarray}

The discord vanishes only where $\eta$ does. Zero discord under measurement on
qubit 2 means $\rho$ is invariant under
$\sum_{\pm}(\Pi_{\pm}\otimes\mathbb{I})\rho(\Pi_{\pm}\otimes\mathbb{I})$ for some
$\hat{n}$. Since $\sum_{\pm}\Pi_{\pm}\sigma_{i}\Pi_{\pm}=n_{i}(\hat{n}\cdot
\vec{\sigma})$, invariance of Eq.~(\ref{eq:bloch}) requires
$c_{i}n_{i}n_{j}=c_{j}\delta_{ij}$, hence $n_{i}^{2}=1$ for every $i$ with
$c_{i}\neq0$: at most one correlation component may survive. Here $c_{1}=-c_{2}$,
so the two transverse components vanish together or not at all, and the condition
collapses to $\eta=0$. That happens only on the edges $p=1$ and $q=1$, where the
one-sided state degenerates to $(\mathbb{I}/2)\otimes|0\rangle\langle0|$ or
becomes diagonal in the computational product basis. Both edges are reached only
asymptotically, so $\mathcal{D}>0$ everywhere in the open square, for any rate
profile. Unlike the thresholds of Eq.~(\ref{eq:boundaries}), the discord has no
boundary in the interior at all.

\subsection{The threshold diagram}
\label{sec:diagram}
\begin{figure}[t]
\includegraphics[width=\columnwidth]{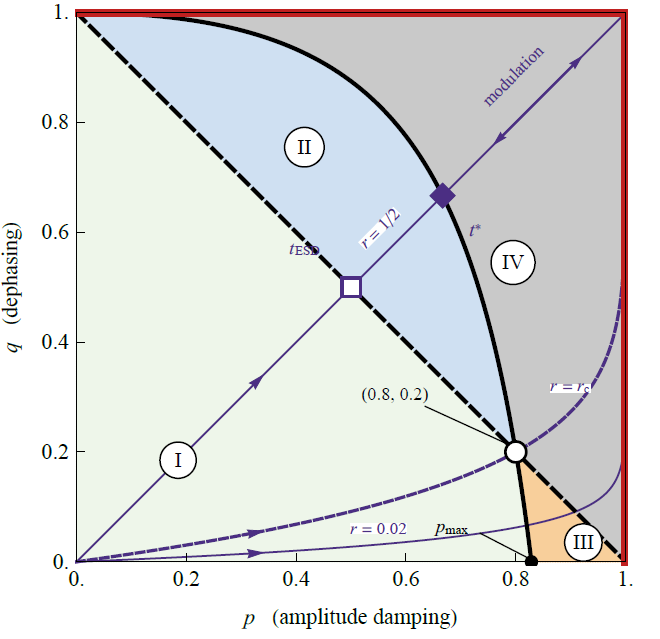}
\caption{Threshold landscape in the $(p,q)$ plane. Solid: the one-sided
advantage boundary $4(1-p)(1-q)=p^{2}$, meeting the axis at
$p_{\max}=2\sqrt{2}-2$. Dashed: the two-sided threshold $p+q=1$, which is
simultaneously sudden death and loss of advantage by Eq.~(\ref{eq:FN}). Red: the
edges $\eta=0$, the only zeros of the one-sided negativity and of the discord.
The two boundaries cross transversally at $(0.8,0.2)$ (circle) and cut the square
into the four regions I--IV described in the text. Thin curves: static
trajectories $q=1-(1-p)^{2r}$ for $r=0.02$, for $r=r_{c}$ (dashed) and for
$r=1/2$, arrows towards increasing $t$. The path depends on $r$ alone; the rate
profile fixes only the motion along it. On the diagonal the markers are
$\gamma_{1,0}t_{\rm ESD}=\ln2$ (square) and $\gamma_{1,0}t^{*}=\ln3$ (diamond),
and the double arrow is the back-and-forth motion produced by modulation.}
\label{fig:diagram}
\end{figure}
Figure~\ref{fig:diagram} collects the results of this section in the plane of the
accumulated damping parameters. Two curves lie in the interior of the unit
square. The first is the locus $2\eta=p$, on which the one-sided fidelity
attains the classical benchmark; it meets the $q=0$ axis at
\begin{equation}
p_{\max}=2\sqrt{2}-2\simeq0.8284 .
\label{eq:pmax}
\end{equation}
The second is the straight line $p+q=1$, on which the two-sided negativity
vanishes and, by Eq.~(\ref{eq:FN}), the two-sided fidelity reaches $2/3$ at the
same instant. Nothing else contributes an interior boundary. The one-sided
negativity and the discord of either state vanish only with $\eta$, hence only on
the edges $p=1$ and $q=1$. Every threshold of the model is therefore one of these
two curves or the boundary of the square, and none of them depends on the
decoherence rates. The rates fix only the trajectory followed within this
landscape.

The two boundaries intersect transversally at
\begin{equation}
(p,q)=(0.8,\,0.2),
\label{eq:intersection}
\end{equation}
the slope of the curve there being $-6$ against $-1$ for the line, and the four
regions so defined exhaust the combinations of entanglement and quantum advantage
accessible to the two geometries. In region~I both resources remain entangled and
beat the benchmark. In region~II the two-sided state is separable and useless
while the one-sided state keeps its advantage. Region~III reverses the ordering:
at equal accumulated damping the one-sided state is still entangled but no longer
useful, while the two-sided state exceeds the benchmark. In region~IV both
criteria fail and the two-sided state is separable as well. Regions III and IV
together are the one-sided useless-entanglement region, and their disappearance
in the symmetric geometry is what identifies that region as a consequence of
asymmetry.

\section{Trajectories through the landscape}
\label{sec:dynamics}

Suppose the two rates share a modulation, so that
$\gamma_{\phi}(t)=r\,\gamma_{1}(t)$ with $r$ constant. Then
$\Gamma_{\phi}=r\,\Gamma_{1}$ at every instant, and Eq.~(\ref{eq:pq}) gives
\begin{equation}
1-q=(1-p)^{2r}.
\label{eq:path}
\end{equation}
The path traced in Fig.~\ref{fig:diagram} is fixed by $r$ alone. Static,
modulated or fast, the state follows the same curve; what the rate profile
decides is how fast it moves along that curve, and whether the motion is always
forward. The thin curves in Fig.~\ref{fig:diagram} are three members of the
family. We take the two questions in turn, first where a fixed path meets the
fixed boundaries, then what modulation does to the motion along it.

\subsection{Static channels and the critical ratio}
\label{sec:static}

Write $x\equiv e^{-\Gamma_{1}(t)}$, so that $p=1-x$, $q=1-x^{2r}$ and
$\eta=x^{(1+2r)/2}$. The two interior boundaries of Fig.~\ref{fig:diagram} are met
at
\begin{equation}
x+x^{2r}=1
\qquad\text{and}\qquad
2x^{(1+2r)/2}=1-x ,
\label{eq:crossings}
\end{equation}
respectively. For static rates $\Gamma_{1}$ increases monotonically, each
equation has one root, and each boundary is crossed once and never recovered.

Which comes first depends only on $r$. The two boundaries meet at $(0.8,0.2)$,
Eq.~(\ref{eq:intersection}), and the path~(\ref{eq:path}) runs through that point
when $(1/5)^{2r}=4/5$, that is when
\begin{equation}
r_{c}=\frac{\ln(5/4)}{2\ln5}\simeq0.0693 .
\label{eq:rc}
\end{equation}
This is the dashed thin curve in Fig.~\ref{fig:diagram}, the one member of the
family that reaches both boundaries at once. Paths with $r>r_{c}$ pass above the
intersection and enter region~II, where the two-sided pair has disentangled while
the one-sided pair still beats the benchmark. Paths with $r<r_{c}$ pass below,
into region~III, and the ordering reverses. Since $r_{c}$ is small, a link would
need dephasing some fourteen times slower than its energy relaxation to sit on
the lower branch. For any real device it is the two-sided resource that fails
first.

From here on $r=1/2$, with both rates sharing $\xi$ and $\Omega$. Then
$q(t)=p(t)$ and $\eta(t)=1-p(t)$ exactly, the path is the diagonal of
Fig.~\ref{fig:diagram}, and Eq.~(\ref{eq:crossings}) reduces to $2x=1$ and $3x=1$.
For $\xi=0$,
\begin{equation}
\gamma_{1,0}t_{\rm ESD}=\ln2\simeq0.693,
\qquad
\gamma_{1,0}t^{*}=\ln3\simeq1.099,
\label{eq:statictimes}
\end{equation}
the square and the diamond on that diagonal. The one-sided
useless-entanglement regime is then the half-line $t>t^{*}$. Because
$\Delta\Gamma_{k}\to0$ as $\Omega\to\infty$, the fast-modulated channel
reproduces Eq.~(\ref{eq:statictimes}) to $\mathcal{O}(\gamma_{1,0}/\Omega)$, and
its curves sit on top of the static ones in Figs.~\ref{fig:onesided} and
\ref{fig:twosided}.

\subsection{Modulated channels: revivals and sudden birth}
\label{sec:modulated}

For $\xi=3$ and $\Omega=2.5\gamma_{1,0}$ the accumulated rate turns over at
$\gamma_{1,0}t\simeq0.76$ and recovers until $\gamma_{1,0}t\simeq1.75$, with
$\Delta\Gamma_{1}=1.278$ from Eq.~(\ref{eq:backflow}). Over that interval the
coherence factor is restored by $e^{\Delta\Gamma_{1}}\simeq3.59$. The state runs
backwards along the diagonal of Fig.~\ref{fig:diagram}, the double arrow there,
and both boundaries can be re-crossed.

Figure~\ref{fig:onesided} shows what this does under one-sided exposure. The
useless-entanglement regime is no longer a half-line but a sequence of finite
windows, separated by intervals in which the fidelity climbs back above $2/3$ and
the pair is worth using again. The negativity in the upper panel never reaches
zero, as Eq.~(\ref{eq:neg1s}) requires. What modulation restores here is the
usefulness of an entanglement that was never lost.

Two-sided exposure is another matter, and Fig.~\ref{fig:twosided} shows it. The
same reversal drives $p+q$ back below unity, so the entanglement does not merely
recover but reappears after having strictly vanished. The static channel crosses
Eq.~(\ref{eq:esd}) once, at $t_{\rm ESD}=\ln2$, and stays dead. The modulated
channel runs a death-birth-death cycle at $\gamma_{1,0}t\simeq0.178$, $1.599$ and
$1.893$, with peak revived negativity $\mathcal{N}_{2s}\simeq0.0423$, magnified
in the inset. This is entanglement sudden birth in the sense of
Ref.~\cite{Lopez2008}, produced by information returning from the environment as
in Refs.~\cite{Bellomo2007,Xu2010,LiuBH2011}.

The revival is small, and Sec.~\ref{sec:backflow} says why it must be. The fold
in $\Gamma_{k}$ never exceeds the drift over one period, Eq.~(\ref{eq:bound}), so
the trajectory cannot be pushed back past where it stood one modulation period
earlier. No threshold is re-crossed more than once. There is at most one window
of restored advantage per boundary and at most one sudden birth, at any $\Omega$
and any admissible depth; raising $\xi$ towards $\xi_{\max}$ deepens the single
fold and does nothing else.

\subsection{Ordering of resource lifetimes}
\label{sec:lifetimes}

The lower panel of Fig.~\ref{fig:twosided} puts the static two-sided lifetimes on
a common scale, each normalised to unity at $t=0$. Negativity and fidelity margin
coincide for $t<t_{\rm ESD}$, which is Eq.~(\ref{eq:FN}) and not a coincidence of
scaling: on that interval they are the same function. Past $t_{\rm ESD}$ they
separate. The negativity is clamped at zero while $\bar{F}^{2s}$ keeps falling.
With $x=e^{-\gamma_{1,0}t}$ one has $\bar{F}^{2s}=(2-x+2x^{2})/3$, whose exact
minimum is
\begin{equation}
\bar{F}^{2s}_{\min}=\frac{5}{8}
\qquad\text{at}\qquad
\gamma_{1,0}t=\ln4 ,
\label{eq:Fmin}
\end{equation}
after which it climbs back towards $2/3$ from below as the resource degenerates
into $|00\rangle\langle00|$. The benchmark is approached but never regained, since
a product state teleports nothing.

The discord decays only asymptotically and stays strictly positive throughout,
sudden death included \cite{Werlang2009,Mazzola2010}. At $t_{\rm ESD}$ it still
holds $\mathcal{D}\simeq0.075$ of its initial value, the triangle in
Fig.~\ref{fig:twosided}. Over a finite interval, then, three statements about the
same distributed pair are true together: it is separable, it is useless for
teleportation, and it is not classical. The first two coincide exactly, by
Eq.~(\ref{eq:FN}). The third is independent of both, and by
Sec.~\ref{sec:discord} it fails nowhere in the open square.

\section{Conclusion}
\label{sec:conclusion}

Once $p(t)$ and $q(t)$ are recognised as the only time-dependent inputs, a
time-varying amplitude- and phase-damping link stops being a dynamical problem.
What is left is a fixed landscape in the unit square and a trajectory that the
environment selects. The landscape supplies one fidelity formula for both
exposure geometries, the sudden-death line $p+q=1$, the fidelity-negativity
relation that holds up to it, and the critical ratio $r_{c}\simeq0.0693$ below
which the ordering of the two lifetimes reverses. Two of its features are worth
isolating. One-sided exposure has no interior entanglement boundary at all, so
the pair disentangles only asymptotically however the rates vary. Neither does
the discord, which leaves an interval in which the distributed state is
separable, below the classical benchmark, and still nonclassical.

The trajectory is constrained by complete positivity alone. For harmonic
modulation the constraint is tight, and the three statements of
Sec.~\ref{sec:backflow} collapse into one: the backflow, the peak that precedes
it, and the admissible depth $\xi_{k}\leq4.6033$ are the same condition. Because
the fold never exceeds the drift over one period, no threshold is re-crossed more
than once, and every revival is a single window, entanglement sudden birth
included. A sharper waveform depositing in a small fraction of the period can
return more than $\gamma_{k,0}T_{\rm mod}$; what no waveform can do is return
more than has already been accumulated. The useless-entanglement region belongs
to the landscape rather than the trajectory, and its closure under symmetric
two-sided noise marks it as an artifact of asymmetry, not of combined damping.

Two extensions follow from the same picture. Ornstein-Uhlenbeck or
random-telegraph rates in place of Eq.~(\ref{eq:rate}) replace the single
trajectory by an ensemble crossing the same fixed boundaries, which turns the
questions asked here into questions of outage probability and quantum-advantage
duty cycle. Letting the two arms differ statistically breaks the symmetry
underlying Eq.~(\ref{eq:FN}) while leaving the fidelity formula intact. The
useless-entanglement gap should then reopen between the two boundaries of
Fig.~\ref{fig:diagram}, and how wide it becomes is the natural thing to compute
next.



\begin{thebibliography}{99}

\bibitem{Bennett1993} C.~H. Bennett, G. Brassard, C. Cr\'epeau, R. Jozsa,
A. Peres, and W.~K. Wootters, Phys. Rev. Lett. \textbf{70}, 1895 (1993).

\bibitem{Kimble2008} H.~J. Kimble, Nature \textbf{453}, 1023 (2008).

\bibitem{Wehner2018} S. Wehner, D. Elkouss, and R. Hanson, Science \textbf{362},
eaam9288 (2018).

\bibitem{Azuma2023} K. Azuma, S.~E. Economou, D. Elkouss, P. Hilaire, L. Jiang,
H.-K. Lo, and I. Tzitrin, Rev. Mod. Phys. \textbf{95}, 045006 (2023).

\bibitem{Hu2023} X.-M. Hu, Y. Guo, B.-H. Liu, C.-F. Li, and G.-C. Guo,
Nat. Rev. Phys. \textbf{5}, 339 (2023).

\bibitem{Qiu2025} J. Qiu, Y. Liu, L. Hu, Y. Wu, J. Niu, L. Zhang, W. Huang,
Y. Chen, J. Li, S. Liu, Y. Zhong, L. Duan, and D. Yu, Sci. Bull. \textbf{70},
351 (2025).

\bibitem{Chou2018} K.~S. Chou, J.~Z. Blumoff, C.~S. Wang, P.~C. Reinhold,
C.~J. Axline, Y.~Y. Gao, L. Frunzio, M.~H. Devoret, L. Jiang, and
R.~J. Schoelkopf, Nature \textbf{561}, 368 (2018).

\bibitem{Main2025} D. Main, P. Drmota, D.~P. Nadlinger, E.~M. Ainley,
A. Agrawal, B.~C. Nichol, R. Srinivas, G. Araneda, and D.~M. Lucas, Nature
\textbf{638}, 383 (2025).

\bibitem{Hermans2022} S.~L.~N. Hermans, M. Pompili, H.~K.~C. Beukers, S. Baier,
J. Borregaard, and R. Hanson, Nature \textbf{605}, 663 (2022).

\bibitem{Knaut2024} C.~M. Knaut, A. Suleymanzade, Y.-C. Wei, D.~R. Assumpcao,
P.-J. Stas, Y.~Q. Huan, B. Machielse, E.~N. Knall, M. Sutula, G. Baranes,
\textit{et al.}, Nature \textbf{629}, 573 (2024).

\bibitem{Wei2025} Y.-C. Wei, P.-J. Stas, A. Suleymanzade, G. Baranes,
F. Machado, Y.~Q. Huan, C.~M. Knaut, S.~W. Ding, M. Merz, E.~N. Knall,
\textit{et al.}, Science \textbf{388}, 509 (2025).

\bibitem{Thomas2024} J.~M. Thomas, F.~I. Yeh, J.~H. Chen, J.~J. Mambretti,
S.~J. Kohlert, G.~S. Kanter, and P. Kumar, Optica \textbf{11}, 1700 (2024).

\bibitem{Liu2024} J.-L. Liu, X.-Y. Luo, Y. Yu, C.-Y. Wang, B. Wang, Y. Hu,
J. Li, M.-Y. Zheng, B. Yao, Z. Yan, \textit{et al.}, Nature \textbf{629}, 579
(2024).

\bibitem{Rad2025} H. Aghaee Rad, T. Ainsworth, R.~N. Alexander, B. Altieri,
M.~F. Askarani, R. Baby, L. Banchi, B.~Q. Baragiola, J.~E. Bourassa,
R.~S. Chadwick, \textit{et al.}, Nature \textbf{638}, 912 (2025).

\bibitem{Ren2017} J.-G. Ren, P. Xu, H.-L. Yong, L. Zhang, S.-K. Liao, J. Yin,
W.-Y. Liu, W.-Q. Cai, M. Yang, L. Li, \textit{et al.}, Nature \textbf{549}, 70
(2017).

\bibitem{Horodecki1999} M. Horodecki, P. Horodecki, and R. Horodecki,
Phys. Rev. A \textbf{60}, 1888 (1999).
\bibitem{RHorodecki1996}
R. Horodecki, M. Horodecki, and P. Horodecki,
Phys. Lett. A \textbf{222}, 21 (1996).
\bibitem{Yu2004} T. Yu and J.~H. Eberly, Phys. Rev. Lett. \textbf{93}, 140404
(2004).

\bibitem{Yu2009} T. Yu and J.~H. Eberly, Science \textbf{323}, 598 (2009).

\bibitem{Almeida2007} M.~P. Almeida, F. de Melo, M. Hor-Meyll, A. Salles,
S.~P. Walborn, P.~H. Souto Ribeiro, and L. Davidovich, Science \textbf{316},
579 (2007).

\bibitem{Laurat2007} J. Laurat, K.~S. Choi, H. Deng, C.~W. Chou, and
H.~J. Kimble, Phys. Rev. Lett. \textbf{99}, 180504 (2007).

\bibitem{Ollivier2001} H. Ollivier and W.~H. Zurek, Phys. Rev. Lett.
\textbf{88}, 017901 (2001).

\bibitem{Henderson2001} L. Henderson and V. Vedral, J. Phys. A: Math. Gen.
\textbf{34}, 6899 (2001).

\bibitem{Modi2012} K. Modi, A. Brodutch, H. Cable, T. Paterek, and V. Vedral,
Rev. Mod. Phys. \textbf{84}, 1655 (2012).

\bibitem{Werlang2009} T. Werlang, S. Souza, F.~F. Fanchini, and
C.~J. Villas Boas, Phys. Rev. A \textbf{80}, 024103 (2009).

\bibitem{Maziero2009} J. Maziero, L.~C. C\'eleri, R.~M. Serra, and V. Vedral,
Phys. Rev. A \textbf{80}, 044102 (2009).

\bibitem{Mazzola2010} L. Mazzola, J. Piilo, and S. Maniscalco,
Phys. Rev. Lett. \textbf{104}, 200401 (2010).

\bibitem{Ali2010} M. Ali, A.~R.~P. Rau, and G. Alber, Phys. Rev. A \textbf{81},
042105 (2010).

\bibitem{Chen2011} Q. Chen, C. Zhang, S. Yu, X.~X. Yi, and C.~H. Oh,
Phys. Rev. A \textbf{84}, 042313 (2011).

\bibitem{Klimov2018} P.~V. Klimov, J. Kelly, Z. Chen, M. Neeley, A. Megrant,
B. Burkett, R. Barends, K. Arya, B. Chiaro, Y. Chen, \textit{et al.},
Phys. Rev. Lett. \textbf{121}, 090502 (2018).

\bibitem{Burnett2019} J.~J. Burnett, A. Bengtsson, M. Scigliuzzo, D. Niepce,
M. Kudra, P. Delsing, and J. Bylander, npj Quantum Inf. \textbf{5}, 54 (2019).

\bibitem{Schloer2019} S. Schl\"or, J. Lisenfeld, C. M\"uller, A. Bilmes,
A. Schneider, D.~P. Pappas, A.~V. Ustinov, and M. Weides, Phys. Rev. Lett.
\textbf{123}, 190502 (2019).

\bibitem{Martinez2021} J. Etxezarreta Martinez, P. Fuentes, P. Crespo, and
J. Garcia-Fr\'ias, npj Quantum Inf. \textbf{7}, 115 (2021).

\bibitem{Martinez2023} J. Etxezarreta Martinez, P. Fuentes,
A. deMarti iOlius, J. Garcia-Frias, J.~R. Fonollosa, and P.~M. Crespo,
Phys. Rev. Research \textbf{5}, 033055 (2023).

\bibitem{Vasylyev2016} D. Vasylyev, A.~A. Semenov, and W. Vogel,
Phys. Rev. Lett. \textbf{117}, 090501 (2016).

Phys. Rev. Lett. \textbf{101}, 150402 (2008).

\bibitem{Breuer2009} H.-P. Breuer, E.-M. Laine, and J. Piilo,
Phys. Rev. Lett. \textbf{103}, 210401 (2009).


\bibitem{Rivas2014} \'A. Rivas, S.~F. Huelga, and M.~B. Plenio,
Rep. Prog. Phys. \textbf{77}, 094001 (2014).

\bibitem{Breuer2016} H.-P. Breuer, E.-M. Laine, J. Piilo, and B. Vacchini,
Rev. Mod. Phys. \textbf{88}, 021002 (2016).

\bibitem{Vega2017} I. de Vega and D. Alonso, Rev. Mod. Phys. \textbf{89},
015001 (2017).

\bibitem{Bellomo2007} B. Bellomo, R. Lo Franco, and G. Compagno,
Phys. Rev. Lett. \textbf{99}, 160502 (2007).

\bibitem{Xu2010} J.-S. Xu, C.-F. Li, M. Gong, X.-B. Zou, C.-H. Shi, G. Chen,
and G.-C. Guo, Phys. Rev. Lett. \textbf{104}, 100502 (2010).

\bibitem{LiuBH2011} B.-H. Liu, L. Li, Y.-F. Huang, C.-F. Li, G.-C. Guo,
E.-M. Laine, H.-P. Breuer, and J. Piilo, Nat. Phys. \textbf{7}, 931 (2011).

\bibitem{Laine2014} E.-M. Laine, H.-P. Breuer, and J. Piilo, Sci. Rep.
\textbf{4}, 4620 (2014).

\bibitem{Gaidi2026} S. Gaidi, N.-E. Abouelkhir, A. Slaoui, M. El Falaki, and
R. Ahl Laamara, Eur. Phys. J. Plus \textbf{141}, 207 (2026).

\bibitem{Seida2021} C. Seida, A. El Allati, N. Metwally, and Y. Hassouni,
Eur. Phys. J. D \textbf{75}, 170 (2021).

\bibitem{Yeo2010} Y. Yeo, J.-H. An, and C.~H. Oh, Phys. Rev. A \textbf{82},
032340 (2010).

\bibitem{Wang2023} Y. Wang, S. Xue, H. Song, and M. Jiang, Phys. Rev. A
\textbf{108}, 062406 (2023).

\bibitem{Zhang2024} H. Zhang, X. Han, G. Zhang, L. Li, L. Cheng, J. Wang,
Y. Zhang, Y. Xia, and C. Xia, Sci. Rep. \textbf{14}, 23885 (2024).

\bibitem{LiuZD2020} Z.-D. Liu, Y.-N. Sun, B.-H. Liu, C.-F. Li, G.-C. Guo,
S. Hamedani Raja, H. Lyyra, and J. Piilo, Phys. Rev. A \textbf{102}, 062208
(2020).

\bibitem{Badziag2000} P. Bad\.ziag, M. Horodecki, P. Horodecki, and
R. Horodecki, Phys. Rev. A \textbf{62}, 012311 (2000).

\bibitem{Verstraete2003} F. Verstraete and H. Verschelde, Phys. Rev. Lett.
\textbf{90}, 097901 (2003).

\bibitem{Peres1996} A. Peres, Phys. Rev. Lett. \textbf{77}, 1413 (1996).

\bibitem{Vidal2002} G. Vidal and R. F. Werner, Phys. Rev. A \textbf{65}, 032314
(2002).

\bibitem{Hall2014} M. J. W. Hall, J. D. Cresser, L. Li, and E. Andersson,
Phys. Rev. A \textbf{89}, 042120 (2014).
\bibitem{Nielsen2010} M. A. Nielsen and I. L. Chuang, \textit{Quantum
Computation and Quantum Information}, 10th Anniversary ed. (Cambridge University
Press, Cambridge, 2010), Chap.~8.

\bibitem{Breuer2002} H.-P. Breuer and F. Petruccione, \textit{The Theory of Open
Quantum Systems} (Oxford University Press, Oxford, 2002).

\bibitem{Andersson2007} E. Andersson, J. D. Cresser, and M. J. W. Hall,
J. Mod. Opt. \textbf{54}, 1695 (2007).
\bibitem{Kofman2001} A. G. Kofman and G. Kurizki, Phys. Rev. Lett. \textbf{87},
270405 (2001).

\bibitem{Amati2024} G. Amati, Phys. Rev. A \textbf{109}, 052433 (2024).

\bibitem{Follia2026} P. M. Follia, B. Vacchini, and H.-P. Breuer, Phys. Rev. A
\textbf{113}, 022203 (2026).

\bibitem{Lopez2008} C. E. L\'opez, G. Romero, F. Lastra, E. Solano, and
J. C. Retamal, Phys. Rev. Lett. \textbf{101}, 080503 (2008).



\end{thebibliography}
\end{document}